%% file: laser.tex
\documentclass[notitlepage]{revtex4-1}
\pdfoutput=1
\usepackage{amsmath}
\usepackage{amssymb}
\usepackage{stmaryrd}
\usepackage{mathrsfs}
\usepackage[utf8]{inputenc}
\usepackage[export]{adjustbox}
\usepackage[dvipsnames]{xcolor}
\usepackage{fancyhdr}
\usepackage{lineno}
\usepackage[english]{babel}
\usepackage{amsmath,amssymb}
\usepackage{siunitx}
\usepackage{soul}
\usepackage{array}
\usepackage{changes}
\begin{document}

\author{Amirhassan Shams-Ansari}
\thanks{These authors contributed equally to this work}
\affiliation{John A. Paulson School of Engineering and Applied Sciences, Harvard University, Cambridge, Massachusetts 02138, USA}

\author{Guanhao Huang}
\thanks{These authors contributed equally to this work}
\affiliation{Institute of Physics, Swiss Federal Institute of Technology Lausanne (EPFL), CH-1015 Lausanne, Switzerland}
\author{Lingyan He}

\affiliation{HyperLight, 501 Massachusetts Avenue, Cambridge, MA 02139}
\author{Zihan Li}
\affiliation{Institute of Physics, Swiss Federal Institute of Technology Lausanne (EPFL), CH-1015 Lausanne, Switzerland}
\author{Jeffrey Holzgrafe}
\affiliation{John A. Paulson School of Engineering and Applied Sciences, Harvard University, Cambridge, Massachusetts 02138, USA}
\author{Marc Jankowski}
\affiliation{Physics \& Informatics Laboratories, NTT Research, Inc., 940 Stewart Drive, Sunnyvale, California 94085, USA}
\affiliation{E. L. Ginzton Laboratory, Stanford University, 348 Via Pueblo Mall, Stanford, California 94305, USA}
\author{Mikhail Churaev}
\affiliation{Institute of Physics, Swiss Federal Institute of Technology Lausanne (EPFL), CH-1015 Lausanne, Switzerland}

\author{Prashanta Kharel}
\affiliation{HyperLight, 501 Massachusetts Avenue, Cambridge, MA 02139}

\author{Rebecca Cheng}
\affiliation{John A. Paulson School of Engineering and Applied Sciences, Harvard University, Cambridge, Massachusetts 02138, USA}

\author{Di Zhu}
\affiliation{John A. Paulson School of Engineering and Applied Sciences, Harvard University, Cambridge, Massachusetts 02138, USA}

\author{Neil~Sinclair}
\affiliation{John A. Paulson School of Engineering and Applied Sciences, Harvard University, Cambridge, Massachusetts 02138, USA}

\author{Boris Desiatov}
\affiliation{John A. Paulson School of Engineering and Applied Sciences, Harvard University, Cambridge, Massachusetts 02138, USA}
\author{Mian Zhang}
\affiliation{HyperLight, 501 Massachusetts Avenue, Cambridge, MA 02139}
\author{Tobias J. Kippenberg}\email{tobias.kippenberg@epfl.ch}
\affiliation{Institute of Physics, Swiss Federal Institute of Technology Lausanne (EPFL), CH-1015 Lausanne, Switzerland}

\author{Marko Lon\v{c}ar}\email{loncar@seas.harvard.edu}
\affiliation{John A. Paulson School of Engineering and Applied Sciences, Harvard University, Cambridge, Massachusetts 02138, USA}
\date{\today}

\title{Reduced Material Loss in Thin-film Lithium Niobate Waveguides}

\begin{abstract}
Thin-film lithium niobate has shown promise for scalable applications ranging from single-photon sources to high-bandwidth data communication systems.
Realization of the next generation high-performance classical and quantum devices, however, requires much lower optical losses than the current state of the art ($\sim$10 milion). Unfortunately, material limitations of ion-sliced thin film lithium niobate have not been explored, and therefore it is unclear how high quality factor  can be achieved in this platform. Here we evaluate the material limited quality factor of thin film lithium niobate photonic platform can be as high as \boldmath$Q\approx 1.8\times10^{8}$ at telecommunication wavelengths, corresponding to a propagation loss of {\SI{0.2}{dB/m}.}
\end{abstract}
\maketitle
Thin-film lithium niobate (TFLN) platform has  enabled a myriad of classical and quantum applications \cite{zhu2021integrated}, many of which crucially rely on low optical loss. 
For instance, the bandwidth of electro-optic (EO) frequency combs \cite{zhang2019broadband} and the efficiency of microwave-to-optical transducers
\cite{Holzgrafe:20,McKenna:20} are proportional to the resonator quality factor $Q$ or (loss rate)\textsuperscript{-1}. 
Currently the lowest-reported optical loss in ion-sliced TFLN waveguides is $\sim$ 3 dB/m \cite{zhang2017monolithic}, which compares favorably to many photonic platforms. 
At the same time, a loss of $\sim$ 0.2 dB/m
was measured using whispering gallery mode resonators created by polishing bulk congruent LN \cite{ilchenko2004nonlinear}. 
It is currently an open question if TFLN can reach this, and ideally even lower, loss rates. For example, it has been speculated that ion slicing process,  used to create TFLN from bulk LN \cite{rabiei2004optical}, may result in implantation damage that could yield higher optical absorption in TFLN than in polished bulk LN.

Here, we first develop a post-fabrication process based on annealing in  O\textsubscript{2} atmosphere in order to reduce material absorption rate $\kappa_\mathrm{abs}$ of TFLN platform, and then we use  Kerr-calibrated linear response measurements \cite{liu2020high} to evaluate $\kappa_\mathrm{abs}$.   This requires parameters such as the nonlinear refractive index $n_2$ as well as the ratio between the photothermal and Kerr-induced cross-phase modulation (XPM) responses (at bandwidths $<$10 MHz) $\gamma=\chi_\mathrm{therm}/\chi_\mathrm{Kerr}$ of TFLN. 
These, and other parameters, are determined by performing laser pump-probe measurements on timescales shorter than the response time of deleterious photorefractive (PR) effects in LN \cite{jiang2017fast}. Using these pump-probe techniques, we determine that the material limited loss in ion-sliced LN is $\sim$1.5 dB/m, and demonstrate an annealing process that reduces this to $\sim$0.2 dB/m that approaches the limit of the bulk LN.

\begin{figure}[ht!]
\vspace{-1ex}
\centering\includegraphics[width=0.47\textwidth, page=2]{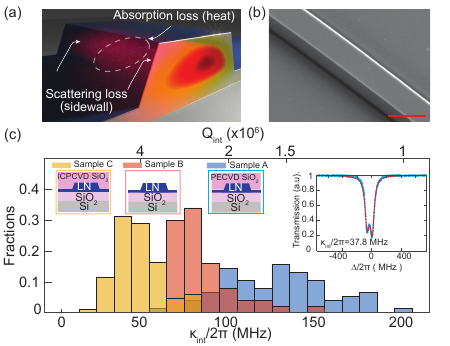}
\vspace{-2ex}
\caption{Low material loss in TFLN micro-ring resonators using thermal annealing. \textbf{(a)} Conceptual representation of optical loss in a waveguide at low powers. Energy from an optical mode (profile cross-section shown) is dissipated due to photothermal absorption and sidewall roughness-induced losses, which can result in generation of heat and scattering of light, respectively. 
\textbf{(b)} A scanning electron microscope image of a fabricated waveguide. Scale bar represents \SI{1}{\micro m}. \textbf{(c)} Statistics of intrinsic loss rates ($Q$-factors) for resonators, denoted as Samples A-C, created by different fabrication methods, as discussed in the main text. Vertical axis represents the fraction of the total number of resonances measured for the TE mode in each resonator for wavelengths between 1480 nm and 1680 nm. Left insets: Schematized wafer cross-sections for Samples A-C. Right inset: A split resonance observed from an annealed resonator, with extracted intrinsic loss rate. Horizontal axis denotes laser detuning from resonance.}
\label{fig:1}
\vspace{-2ex}
\end{figure}

The micro-rings are fabricated on a \SI{600}{nm}-thick x-cut LN thin-film bonded to a \SI{4.7}{\micro m}-thick layer of thermal oxide on a silicon wafer (NanoLN). 
Electron-beam lithography followed by physical reactive Ar\textsuperscript{+} ion etching, with target etch-depth of 300 nm, yields micro-ring resonators of \SI{140}{\micro m} radius and a waveguide top-width of \SI{2.4}{\micro m} (Fig \ref{fig:1}b). 

We prepare three sets of resonators from this wafer in the same fabrication run: cladded (Sample-A), annealed (Sample-B), and annealed-cladded-annealed (Sample-C) resonators. 
Sample-A is fabricated with the process reported in Ref. \cite{zhang2017monolithic}.
That is, the resonators are cladded with an 800~nm-thick layer of SiO\textsubscript{2} using plasma-enhanced chemical vapor deposition with substrate temperature of 300\textsuperscript{$\circ$}C. 
For Samples-B and -C, the resonators are annealed at atmospheric pressures in O\textsubscript{2} at 520\textsuperscript{$\circ$}C for two hours. 
The annealing step is used to improve the crystallinity of TFLN, thereby repairing potential damages \cite{leidinger2016influence} caused by ion slicing \cite{han2015optical}. 
We clad Sample-C with a 800 nm-thick layer of SiO\textsubscript{2} deposited using inductively coupled plasma chemical vapour deposition (ICPCVD) at 80\textsuperscript{$\circ$}C, and then re-anneal it under the same conditions.We emphasize the low temperature nature (80C) of ICPCVD process, which we found to be  important to maintain the benefits of the annealing step.
We measure a mean $Q_{\mathrm{int}}$ of $1.5$, $2.5$, and $5$ million in Samples-A,-B, and -C, respectively (Fig.~\ref{fig:1}c). We note that all resonances from Samples -B and -C exhibit asymmetric mode splittings due to Rayleigh back-scattering (Fig.~\ref{fig:1}c).

\begin{figure}[ht!]
\centering\includegraphics[width=0.45\textwidth, page=2]{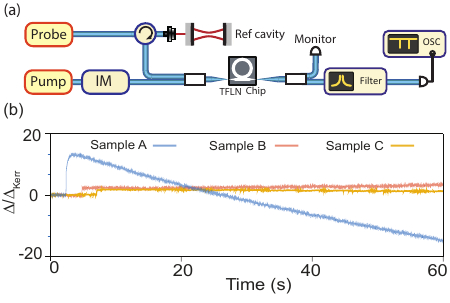}
\vspace{-3ex}
\caption{\textbf{a} The laser pump-probe setup for a step-response measurement to characterize the timescale of photorefractive effects. \textbf{b} Dynamics of step excitation-induced cavity resonance shift (normalized to Kerr shift) of the probe after switching-off the pump. Measurement is referenced to a thermally stabilized Fabry-Perot cavity. Pump is switched off at slightly different times for Samples A-C. Timing resolution is 1 ms. IM: Intensity Modulator, OSC: Oscilloscope} \label{fig:2}
\vspace{-2ex}
\end{figure}

Before the Kerr-calibrated response measurements, we first determine the timescale of PR effects in the micro-rings since PR-induced resonance frequency change could distort the inferred material response at low modulation frequencies. \cite{jiang2017fast}.
To do this, we optically pump a micro-ring then, after extinguishing the pump, we repeatedly measure one of its resonances  with a probe, monitoring the time-dependence of the detuning of the resonance (Fig.~\ref{fig:2}a). 
The detuning is normalized to the Kerr shift (discussed later) for convenient comparison.
For Sample-A, we observe a blue shift with a time constant of $\sim\SI{100}{s}$, indicative of PR effects.
We did not observe PR behavior for Samples-B and -C over time scales of up to a minute (Fig.~\ref{fig:2}b).
Thus, the PR effect can be ignored for measurements at timescales significantly shorter than 100 s (bandwidths $>>$ 0.01 Hz).

To calibrate the absorption rate of different devices, we need to evaluate
\begin{equation}
	\kappa_\mathrm{abs} = \frac{2c\nu n_2\gamma}{n_gn_\mathrm{eff}Vd\nu/dP_\mathrm{abs}},
\end{equation}
where $V$ the optical mode volume, $n_g$ the group index, $n_\mathrm{eff}$ the effective index, and $d\nu/dP_\mathrm{abs}$ the photothermal frequency shift gradient at pump frequency $\nu$, mainly determined by the material thermo-optic coefficients~\cite{moretti2005temperature} and is determined from simulation, see Supplementary materials. The response ratio $\gamma$ is the DC offset of the measured response function $\gamma(\omega)$ in Fourier domain, and is obtained through fitting. The material absorption rate also requires a good knowledge of $n_2$ from TFLN. We did an auxiliary pump-probe measurement to obtain $n_2 = \SI{1.51e-19}{m^2W^{-1}}$ for our TFLN, details see Supplementary Material.  
We did not employ the commonly-used thermal triangle technique\cite{gao2021probing} to determine $n_2$ to avoid PR effects.

\begin{figure}[ht!]
\centering\includegraphics[width=0.45\textwidth, page=1]{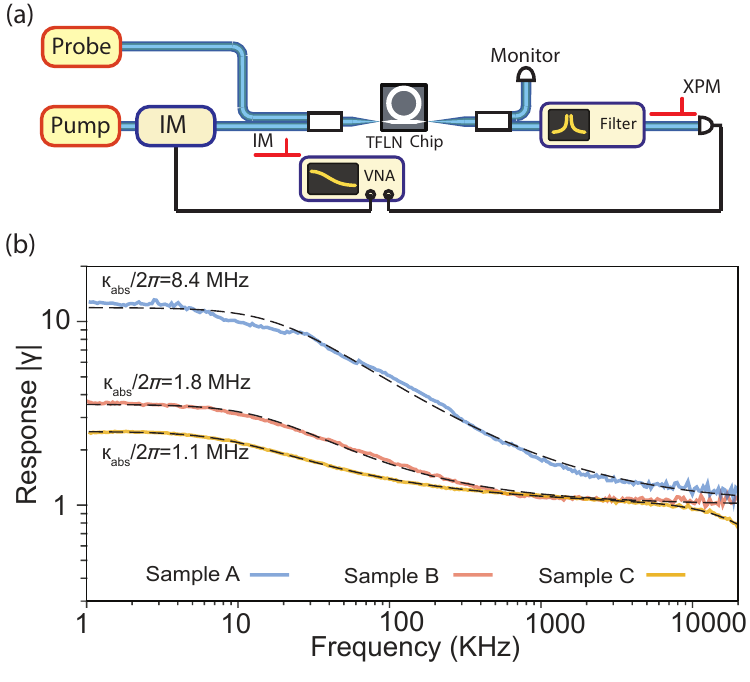}
\vspace{-2ex}
\caption{\textbf{(a)} Laser pump-probe setup for measuring the DC offset of the measured response function ($\gamma$) with calibrated modulation tones. \textbf{(b)} Measured photothermal and XPM responses for Samples A-C. A fit to the curve determines $\gamma$.
VNA: Vector Network Analyzer, IM: Intensity Modulation, PM: Phase Modulation, RSA: Real Time Spectrum Analyzer.}\label{fig:4}
\vspace{-2ex}
\end{figure}
The measurement of $\gamma$ is accomplished by modulating an optical pump and measuring the resultant side-of-fringe modulation of a probe, as induced by the material response of TFLN (see Fig.~\ref{fig:4}a).
The pump modulation frequency is varied to elucidate photothermal and Kerr-induced XPM on the probe, which can be distinguished at low ($<$10 kHz) and high ($>$1 MHz) modulation frequencies, respectively \cite{liu2020high}. 
The responses of all the devices were measured, and all yield two plateaus corresponding to either predominantly Kerr induced ($\chi_\mathrm{Kerr}$) or photothermal-induced ($\chi_\mathrm{therm}$) responses.
The ratios of the plateaus $\gamma$ are 11.0, 2.5 and 1.5 for Samples -A, -B, and -C, respectively, indicating that thermal annealing reduces the magnitude of photothermal response. Note that cascaded sum-frequency generation of the pump with the probe will contribute an additional XPM indistinguishable from the Kerr contribution.
Given our system parameters, we expect this change to be $\sim$ 10\%, see Supplementary materials. For all measurements, the wavelength of the pump is $\sim$1550 nm while the probe is at wavelengths detuned several free-spectral ranges of the resonator away.
The pump power is kept low ($<$\SI{1}{mW}) to avoid nonlinearity induced self-feedback.

Finally, using Eq. 1 we calculate $\kappa_\mathrm{abs}/2\pi$ to be 8.4 MHz, 1.8 MHz, and 1.1 MHz for Samples -A, -B, and -C, respectively.
The absorption rate corresponding to each process is calibrated individually, with Sample-C yielding a material-limited quality factor of \SI{180}{Million} (~\SI{0.2}{dB/m}) which is among the highest within integrated photonic platforms, see Table~\ref{tb:1}. 
Our results suggest that the main source of loss in our high-confinement LN waveguides is line-edge roughness-induced scattering, which limits the average intrinsic quality factor $\kappa_\mathrm{int}/2\pi$ of resonances around \SI{1550}{nm} to 
\SI{34}{MHz}.(Fig. \ref{fig:1}c).

\begin{table}[htbp]
\vspace{-3ex}
\centering
\caption{\bf Comparison of 
material-limited properties among state-of-the-art integrated photonics platforms \cite{gao2021probing}}
\label{tb:1}
\setlength\extrarowheight{-2pt}
\setlength{\tabcolsep}{2pt}
\begin{tabular}{ccccc}
\hline
Platform & $n$ & $\chi_2$(pm/V)& $n_2$($10^{-20}$\SI{}{m^2/W}) & Q-factor  ($10^6$)\\
\hline
SiO\textsubscript{2} & 1.45 &0& 2.2 & 5363 \\
SiN & 2.0 &0& 24 & 287 \\
Ta\textsubscript{2}O\textsubscript{5}&2.0&0&62&2.64\\
Al\textsubscript{0.2}Ga\textsubscript{0.8}As&3.3&119&2600&2.04\\
\textbf{TFLN}
(this work) &2.2&30&\textbf{15}
& \textbf{180} \\

\hline
\end{tabular}

  \label{tab:shape-functions}
\end{table}

In conclusion, we demonstrated that post-fabrication annealing and low-temperature oxide cladding can significantly reduce optical absorption in TFLN waveguides.
Absorption at telecommunication wavelengths is reduced by removing damage potentially caused by ion implantation and reactive-ion etching even at temperatures in which other chemical bonds (e.g. Si-H and O-H) are still present.
Consequently, annealing reduces the material absorption loss significantly over a broad frequency range \cite{leidinger2016influence}.
Our annealing technique yielded absorption-limited loss on par with $\sim$0.2 dB/m measured in bulk LN \cite{ilchenko2004nonlinear}, corresponding to a material-limited Q-factor of 180 Million. 
We anticipate that, by improving TFLN fabrication strategies, $Q$-factors approaching the material limit can be achieved, reaching the low-loss regime required for transformative quantum and classical technology,
e.g. deterministic room-temperature single photon source with periodically-poled TFLN micro-ring~\cite{lu2020toward}.
\newline
\textbf{Funding:} Air Force Office of Scientific Research (FA9550-19-1-0376), Swiss National Science Foundation (SNSF) (185870, 192293). Defense Advanced Research Projects Agency (HR0011-20-C-0137, HR0011-20-2-0046). 
\newline
\textbf{Acknowledgement:} We acknowledge fruitful discussions with Martin M. Fejer, Linbo Shao, Maodong Gao, and Christian Reimer. Fabrication is performed at the Harvard University Center for Nanoscale Systems (CNS).
\newline
\textbf{Disclosures:} L.H, P.K, M.Z and M.L are involved in developing TFLN technologies at Hyperlight Corporation.
\newline
\textbf{Disclaimer:} The views, opinions and/or findings expressed are those of the author and should not be interpreted as representing the official views or policies of the Department of Defense or the U.S. Government.
\vspace{-2ex}
\vspace{-2ex}

\section*{Supplementary materials}
\input{SI/SI_sub}
\newpage
\bibliographystyle{ieeetr}
\bibliography{laser}

\end{document}

%% file: SI/SI_sub.tex
\subsection*{Estimated contribution of cascaded sum-frequency generation to XPM}
Determining $n_2$ relies on pump-probe measurements (e.g. the detuning of a resonance with modulating the pump) in our experiment. 
In this case, cascaded sum-frequency generation (SFG) of the bright pump with the dim probe will contribute to an effective XPM term since the shift of the cavity resonances is measured as a function of pump power rather than probe power.
However, for TFLN, owing to its large $\chi\textsuperscript{(2)}$, cascaded second order nonlinearities can contribute an $n_2$ with nearly equal and opposite sign to the pure electronic $n_2$ of LN. 
As a result, values of $n_2$ inferred from Z-scans in bulk material or the thermal triangle in ring resonators may differ by an order of magnitude from the real value\cite{Phillips:11,Conti:02}. We may estimate the strength of these contributions for the waveguides under consideration here by finding both the nonlinear coupling and phase-mismatch between the pump, probe, and generated sum-frequency.

The coupled wave equations for three-wave mixing are given (in power normalized units) by
\begin{subequations}
\begin{align}
\partial_z A_3 = -i\kappa_3 A_1 A_2 \exp(i \Delta k z),\label{CWE3}\\
\partial_z A_2 = -i\kappa_2 A_3 A_1^* \exp(-i \Delta k z),\label{CWE2}\\
\partial_z A_1 = -i\kappa_1 A_3 A_2^* \exp(-i \Delta k z),\label{CWE1}
\end{align}
\end{subequations}
where $\kappa_3/\omega_3 = \kappa_2/\omega_2 = \kappa_1/\omega_1$. In the undepleted limit, we may assume that the bright pump is given by $A_2(z) = A_2(0)\exp(-i \phi_\mathrm{NL,2}(z))$, and $A_1(z) = A_1(0)\exp(-i \phi_\mathrm{NL,1}(z))$, where $\phi_\mathrm{NL,1}(z) = \delta k_{1} z$ is assumed to be approximately linear in $z$. We assume $\delta k_{1(2)} \ll \Delta k$, in which case the sum frequency is given by
\begin{equation}
A_3(z) = -\frac{\kappa_3}{\Delta k} A_1(0) A_2(0)\left(\exp(i (\Delta k + \delta k_1 + \delta k_2) z)-1\right)\label{SF}.
\end{equation}
Substituting Eqn \ref{SF} into Eqns \ref{CWE2}-\ref{CWE3}, we have
\begin{subequations}
\begin{align}
\partial_z A_2 = i\frac{\kappa_2\kappa_3}{\Delta k} |A_1(0)|^2 \exp(-i\delta k_2 z),\label{dA2}\\
\partial_z A_1 = i\frac{\kappa_1\kappa_3}{\Delta k} |A_2(0)|^2 \exp(-i\delta k_1 z),\label{dA1}
\end{align}
\end{subequations}
where we have ignored oscillatory $\exp(-i\Delta k z)$ terms that do not contribute to the average phase accumulated by $A_1$ and $A_2$. Substituting the approximations $A_2(z) = A_2(0)\exp(-i \phi_\mathrm{NL,2}(z))$, and $A_1(z) = A_1(0)\exp(-i \phi_\mathrm{NL,1}(z))$ into Eqns \ref{dA1}-\ref{dA2}, we find
\begin{subequations}
\begin{align}
\partial_z \phi_2(z) = -\frac{\kappa_2\kappa_3}{\Delta k} |A_1(0)|^2 = \gamma_\mathrm{XPM,2}^{(2)}|A_1(0)|^2,\label{dphi2}\\
\partial_z \phi_1(z) = -\frac{\kappa_1\kappa_3}{\Delta k} |A_2(0)|^2 = \gamma_\mathrm{XPM,1}^{(2)}|A_2(0)|^2.\label{dphi1}
\end{align}
\end{subequations}

The cross-phase modulation that occurs during phase-mismatched SFG between the bright pump and dim probe signal is indistinguishable from the XPM typically encountered in $\chi^{(3)}$ media, since both are linear in pump power, and has the opposite sign for typical values of $\Delta k$. The $\Delta k^{-1}$ scaling associated with cascaded nonlinearities may render the strength of this effect a strong function of waveguide geometry. We also note here that while Eqns \ref{dphi1}-\ref{dphi2} are accurate for TM modes in Z-cut thin films, these expressions need to be corrected for TE modes in X-cut thin films. In this case, $\kappa_3(z)$ oscillates periodically during propagation around a ring since the $d_{ijk}$ tensor associated with $\chi^{(2)}$ interaction is not invariant with respect to rotations around the crystalline X axis. We further note that typical waveguide geometries in LN exhibit avoided crossings between TE and TM modes for propagation angles offset from the crystalline axes. For large radii rings, these avoided crossings cause a TE-polarized mode to undergo adiabatic conversion between TE and TM, thereby contributing more rapid oscillations to both $\kappa_3(z)$ and $\Delta k(z)$. In this more general case, we may Fourier series expand $\kappa_3(z)\exp(i\Delta k(z) z)$ and $\kappa_1(z)\exp(-i\Delta k(z) z)$, and solve Eqns \ref{dphi1}-\ref{dphi2} for each Fourier component. The total contribution to the XPM coefficient is given by

\begin{equation}
\gamma_\mathrm{XPM,2}^{(2)} = -\sum_m \frac{\kappa_{3,m}^2\omega_1}{\omega_3\Delta k_m},
\end{equation}
where $\kappa_{3,m}$ is the $m$th Fourier component of $\kappa_3(z)\exp(i(\Delta k(z)-\Delta k(0)) z)$ and $\Delta k_m = k_3-k_2-k_1-m/R$ is the phase-mismatch associated with each Fourier component. For large radii rings $\Delta k_m\approx k_3-k_2-k_1$, and the total contribution from each component may be evaluated using Parseval's theorem. Using this approximation and the waveguide geometry shown in main text Fig.1, we estimate that the contribution of cascaded nonlinearities to the net XPM coefficient is below 10\%.

\subsection*{n2 calibration method}

The material absorption rate requires evaluation of $n_2$. Since there is no reliable literature on $n_2$ of TFLN, we perform measurements on an auxiliary z-cut sample with similar waveguide geometry to determine $n_2$. We chose this specific crystal direction to minimize error of the cascaded $\chi_2$ calculation and mitigate the impact of adiabatic evolution of the modes from TE to TM during propagation in x-cut sample due to birefringence. 

\begin{figure}[ht!]
\centering\includegraphics[width=0.45\textwidth, page=3]{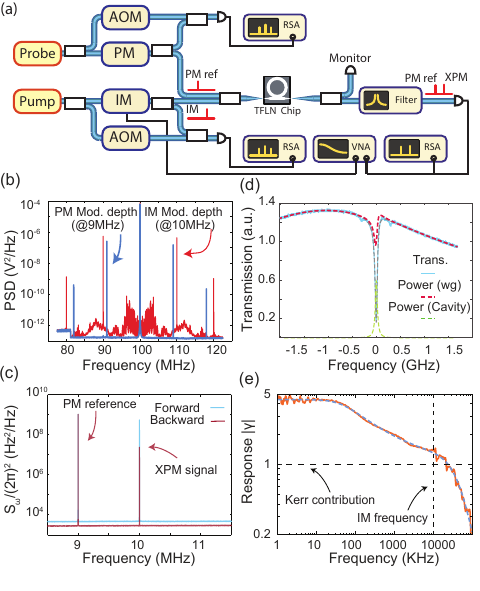}
\vspace{-2ex}
\caption{\textbf{a} Laser pump-probe setup for measuring $n_2$ with calibrated intensity modulation and phase reference tones. 
\textbf{b} Power spectral density of heterodyne measurement at \SI{100}{MHz} offset frequency to calibrate pump intensity modulation and probe phase reference for the XPM-induced modulation. 
\textbf{c} Measured XPM signals at \SI{10}{MHz} along with the phase modulation reference tone at \SI{9}{MHz}. Forward and backward coupling directions are measured to account for facet losses.
\textbf{d} Parameter extraction by fitting a resonance profile of a micro-ring.
\textbf{e} XPM frequency modulation response function. Intensity modulation frequency used to calculate $n_2$ is shown by a vertical dashed line. The horizontal dashed line indicates the pure Kerr XPM response. IM: Intensity Modulation, PM: Phase-Modulation, AOM: Acousto-optic Modulation, RSA: real-time spectrum analyzer, VNA: Vector Network Analyzer} \label{fig:3}
\vspace{-2ex}
\end{figure}

Due to the presence of PR effects, commonly-used thermal triangle technique\cite{gao2021probing} that uses high optical power to determine $n_2$ is not suitable for our platform. Instead, we use a pump-probe scheme (setup see Fig. ~\ref{fig:3}a) similar to the one used in the main text to determine the value of $n_2$ in TFLN at low optical power. The main idea of this measurement is to carefully calibrate how much pump intracavity power density modulation $\delta\rho$ is applied, and how much probe cavity frequency modulation $\delta \nu$ is induced by Kerr effect, after which the value of $n_2$ can be retrieved through relation
\begin{equation}
	\delta\nu = -\frac{2\nu c\overline{n_2}}{\overline{n_gn}}\delta\rho\label{eq:n2_1}
\end{equation}
where $\overline{x}$ are the mode intensity weighted average of the corresponding physical quantities, retrieved from simulation. 

The intracavity power density modulation $\delta\rho$ is determined by the intensity modulation depth $\alpha$ and the waveguide circulating power $P_{\mathrm{WG}}$,
\begin{gather*}
	\delta \rho = \alpha\rho(P_{\mathrm{WG}}).
\end{gather*}
We intensity modulate the pump intensity $I(t) = I_0[1+\alpha \cos(2\pi\Omega_\mathrm{IM} t)]$ at $\Omega_\mathrm{IM}=10$ MHz
and determine the modulation-depth $\alpha$ using heterodyne measurements (Fig. \ref{fig:3}b red spectrum), frequency offset by 100 MHz using Acousto-optic modulators. The intracavity power density $\rho(P_{\mathrm{WG}})$ as a function of waveguide circulating power $P_{\mathrm{WG}}$ depends on many parameters. Apart from cavity coupling rates, the power is also affected by the background etalon formed due to the chip facet reflections. Transmission trace thus consists an overall broad background sine modulation, and a Fano-shaped narrow cavity resonance dip. We express the intracavity power density considering all these effects as
\begin{gather*}
    \rho(P) = \left|\frac{\chi_{\mathrm{cav}}(\Delta)}{1-R(1-\sqrt{\kappa_{\mathrm{ex}}}\chi_{\mathrm{cav}}(\Delta))^2e^{i\theta_{\mathrm{FP}}(\Delta)}}\right|_{\mathrm{extr:}_\Delta}^2\frac{P}{V},\\
    \theta_{\mathrm{FP}}(\Delta) = -2\pi\frac{\Delta}{\Delta\nu_{\mathrm{FP}}} +  \Delta\theta_{\mathrm{FP}},\\
    \chi_{\mathrm{cav}}(\Delta) = \frac{\sqrt{\kappa_{\mathrm{ex}}}}{(\frac{\kappa}{2}+i\Delta)},
\end{gather*}
with $\kappa_{\mathrm{ex}}$ the cavity external coupling rate, $\kappa$ the cavity linewdith, $\Delta$ the laser detuning, $\Delta\nu_{\mathrm{FP}}$ the waveguide background etalon fringe periodicity, $\Delta\theta_{\mathrm{FP}}$ the etalon phase offset, $R$ the characteristic etalon reflectivity, and $V$ the mode volume of the pump mode. Here, power $P$ is the waveguide circulating power at the quadrature point of the waveguide background etalon fringe. All the parameters used in the power density function is fitted from the pump mode transmission profile shown in Fig.\ref{fig:3}d, using fitting function,
\begin{gather*}
    F(\Delta)=\left|\frac{1-\sqrt{\kappa_{\mathrm{ex}}}\chi_{\mathrm{cav}}(\Delta)}{1-R(1-\sqrt{\kappa_{\mathrm{ex}}}\chi_{\mathrm{cav}}(\Delta))^2e^{i\theta_{\mathrm{FP}}(\Delta)}}\right|^2.
\end{gather*}
Sidebands at \SI{300}{MHz} are applied to calibrate the laser detuning. The fitting results are shown in the following table:
\begin{center}
\begin{tabular}{ |c|c|c|c|c| } 
 \hline
 $\kappa_{\mathrm{ex}}/2\pi$ & $\kappa/2\pi$ & $R$ & $\Delta \nu_{\mathrm{FP}}/2\pi$ & $\Delta\theta_{\mathrm{FP}}$\\
 \hline
 \SI{14.2}{MHz} & \SI{47.8}{MHz} & 0.152 & \SI{11.0}{GHz}& \SI{-0.57}{rad}\\ 
 \hline
\end{tabular}
\end{center}

After the intracavity power density function is determined through fitting the cavity transmission trace, we need to calibrate how much Kerr frequency modulation $\delta \nu$ on the probe cavity is induced from a given waveguide circulating power $P$. To calibrate the cavity frequency modulation depths, we use the method from Ref~\cite{gorodetksy2010determination} by comparing the Kerr frequency modulation signal to a reference phase modulation with known depth $\beta$ (calibrated also using heterodyne measurements, Fig. \ref{fig:3}b blue spectrum). The phase modulation $E(t) = E_0e^{i\beta\cos(2\pi\Omega_\mathrm{PM}t)}$ is applied to the probe laser at $\Omega_\mathrm{PM} = \SI{9}{MHz}$, and is visible in Fig. \ref{fig:3}c right next to the cavity frequency modulation signal at $\Omega_\mathrm{IM} = \SI{10}{MHz}$. The reference phase modulation acts as a ruler and allows us to compare and retrieve the cavity frequency modulation depth at different optical powers. To isolate the cavity frequency modulation contributed by Kerr effect from the one from thermal effect, we also measured the XPM response at different pump modulation frequencies using a vector network analyzer. We retrieved the fraction of pure Kerr contribution to the total XPM signal $\Gamma(\Omega_\mathrm{IM})=\chi_\mathrm{Kerr}(\Omega_\mathrm{IM})/\chi_\mathrm{XPM}(\Omega_\mathrm{IM})=0.6$ at $\Omega_\mathrm{IM}=10$ MHz by fitting the measured response (Fig.~\ref{fig:3}e). After that, the Kerr induced cavity frequency modulation can be expressed as
\begin{gather*}
	\delta\nu = \beta\Omega_\mathrm{PM}\Gamma(\Omega_\mathrm{IM})\xi^{1/2}
\end{gather*}
where $\xi=S_{\mathrm{XPM}}/S_{\mathrm{ref}}$ is the power spectral density ratio between the reference phase modulation signal $S_{\mathrm{ref}}$ and the XPM total signal $S_{\mathrm{XPM}}$ measured on the real-time spectrum analyzer. 

Since we do not have direct access to the on-chip waveguide circulating power, and the coupling efficiencies at the chip facets can be different, we mitigate the uncertainties by taking the geometry average of the input power $P_{\mathrm{in}}$ and output power $P_{\mathrm{out}}$ of the chip as the waveguide circulating power $P=\sqrt{P_{\mathrm{in}}P_{\mathrm{out}}}$, measured at the etalon quadrature point. We measure the XPM ratio $\xi_F$ at the given setting, and repeat the measurement (measure ratio $\xi_R$) after reversing the input and output of the micro-ring, in order to take into account different coupling efficiencies at the chip facets. The spectrum when measuring both $\xi_F$ and $\xi_R$ are shown in Fig.~\ref{fig:3}c, and we take their geometric average as well $\xi = \sqrt{\xi_F\xi_R}$.

For all measurements, the wavelength of the pump is $\sim$1550 nm while the probe is at wavelengths detuned several free-spectral ranges of the resonator away.
The pump power is kept low ($<$\SI{1}{mW}) to avoid nonlinearity induced self-feedback. With all relevant parameters measured/fitted, by inverting Eq.\ref{eq:n2_1}, our measurements allow calibrating $n_2$ using
\begin{equation}
	\overline{n_2} = \frac{\beta\Omega_\mathrm{PM}\Gamma(\Omega_\mathrm{IM})[\xi_F\xi_R]^{1/4}\overline{n_g n}}{2c\nu\alpha\rho([P_\mathrm{in}P_\mathrm{out}]^{1/2})},\label{eq:n2}
\end{equation}
and we find a material nonlinear refractive index of $n_2 = \SI{1.51e-19}{m^2W^{-1}}$ for our TFLN. 

Note that cascaded sum-frequency generation of the pump with the probe will contribute an additional XPM indistinguishable from the Kerr contribution. 
Given our system parameters, we expect this change to be $\sim$ 10\%.

All the physical quantities measured/fitted for the $n_2$ calibration are shown in the following table:
\begin{center}
\begin{tabular}{ cc} 
 \hline\hline
 parameters & values \\
 \hline
 $\alpha$ & 0.1619 \\ 
 $\beta$ & 0.1245 \\
 $\Gamma(\Omega_{\mathrm{IM}})$ & 0.60 \\
 $\xi$ & 44.07 \\
  $\xi_R$ & 1.915 \\
  $P_{\mathrm{in}}$ & \SI{680}{\micro W}\\
  $P_{\mathrm{out}}$ & \SI{6.0}{\micro W}\\
 $\rho([P_{\mathrm{in}}P_{\mathrm{out}}]^{1/2})$ & \SI{57.4}{J/m^3}\\
 \hline\hline
\end{tabular}
\end{center}
and the other parameters used in either the $n_2$ calibration or the absorption calibration are retrieved from COMSOL simulation, shown in the following table:
\begin{center}
\begin{tabular}{ cc} 
 \hline\hline
 $n_2$ parameters & values \\
 \hline
 $\overline{n_gn}$ & 4.53\\
 $V$ &\SI{7.47e-16}{m^3}\\
 \hline\hline
 abs parameters & values \\
 \hline
 cladded $V\frac{d\nu}{\nu dP_{\mathrm{abs}}}$ & \SI{2.82e-18}{m^3/W} \\
 uncladded $V\frac{d\nu}{\nu dP_{\mathrm{abs}}}$ & \SI{4.05e-18}{m^3/W}\\
 \hline\hline
\end{tabular}
\end{center}

%% file: laser.bbl
\begin{thebibliography}{10}

\bibitem{zhu2021integrated}
D.~Zhu, L.~Shao, M.~Yu, R.~Cheng, B.~Desiatov, C.~Xin, Y.~Hu, J.~Holzgrafe,
  S.~Ghosh, A.~Shams-Ansari, {\em et~al.}, ``Integrated photonics on thin-film
  lithium niobate,'' {\em Advances in Optics and Photonics}, vol.~13, no.~2,
  pp.~242--352, 2021.

\bibitem{zhang2019broadband}
M.~Zhang, B.~Buscaino, C.~Wang, A.~Shams-Ansari, C.~Reimer, R.~Zhu, J.~M. Kahn,
  and M.~Lon{\v{c}}ar, ``Broadband electro-optic frequency comb generation in a
  lithium niobate microring resonator,'' {\em Nature}, vol.~568, no.~7752,
  pp.~373--377, 2019.

\bibitem{Holzgrafe:20}
J.~Holzgrafe, N.~Sinclair, D.~Zhu, A.~Shams-Ansari, M.~Colangelo, Y.~Hu,
  M.~Zhang, K.~K. Berggren, and M.~Lon\v{c}ar, ``Cavity electro-optics in
  thin-film lithium niobate for efficient microwave-to-optical transduction,''
  {\em Optica}, vol.~7, pp.~1714--1720, Dec 2020.

\bibitem{McKenna:20}
T.~P. McKenna, J.~D. Witmer, R.~N. Patel, W.~Jiang, R.~V. Laer,
  P.~Arrangoiz-Arriola, E.~A. Wollack, J.~F. Herrmann, and A.~H. Safavi-Naeini,
  ``Cryogenic microwave-to-optical conversion using a triply resonant
  lithium-niobate-on-sapphire transducer,'' {\em Optica}, vol.~7,
  pp.~1737--1745, Dec 2020.

\bibitem{zhang2017monolithic}
M.~Zhang, C.~Wang, R.~Cheng, A.~Shams-Ansari, and M.~Lon{\v{c}}ar, ``Monolithic
  ultra-high-q lithium niobate microring resonator,'' {\em Optica}, vol.~4,
  no.~12, pp.~1536--1537, 2017.

\bibitem{ilchenko2004nonlinear}
V.~S. Ilchenko, A.~A. Savchenkov, A.~B. Matsko, and L.~Maleki, ``Nonlinear
  optics and crystalline whispering gallery mode cavities,'' {\em Physical
  review letters}, vol.~92, no.~4, p.~043903, 2004.

\bibitem{rabiei2004optical}
P.~Rabiei and P.~Gunter, ``Optical and electro-optical properties of
  submicrometer lithium niobate slab waveguides prepared by crystal ion slicing
  and wafer bonding,'' {\em Applied Physics Letters}, vol.~85, no.~20,
  pp.~4603--4605, 2004.

\bibitem{liu2020high}
J.~Liu, G.~Huang, R.~N. Wang, J.~He, A.~S. Raja, T.~Liu, N.~J. Engelsen, and
  T.~J. Kippenberg, ``High-yield, wafer-scale fabrication of ultralow-loss,
  dispersion-engineered silicon nitride photonic circuits,'' {\em Nature
  communications}, vol.~12, no.~1, pp.~1--9, 2021.

\bibitem{jiang2017fast}
H.~Jiang, R.~Luo, H.~Liang, X.~Chen, Y.~Chen, and Q.~Lin, ``Fast response of
  photorefraction in lithium niobate microresonators,'' {\em Optics letters},
  vol.~42, no.~17, pp.~3267--3270, 2017.

\bibitem{leidinger2016influence}
M.~Leidinger, K.~Buse, and I.~Breunig, ``Influence of dry-oxygen-annealing on
  the residual absorption of lithium niobate crystals in the spectral range
  from 500 to 2900 nanometers,'' {\em Optical Materials Express}, vol.~6,
  no.~1, pp.~264--269, 2016.

\bibitem{han2015optical}
H.~Han, L.~Cai, and H.~Hu, ``Optical and structural properties of
  single-crystal lithium niobate thin film,'' {\em Optical Materials}, vol.~42,
  pp.~47--51, 2015.

\bibitem{moretti2005temperature}
L.~Moretti, M.~Iodice, F.~G. Della~Corte, and I.~Rendina, ``Temperature
  dependence of the thermo-optic coefficient of lithium niobate, from 300 to
  515 k in the visible and infrared regions,'' {\em Journal of Applied
  Physics}, vol.~98, no.~3, p.~036101, 2005.

\bibitem{gao2021probing}
M.~Gao, Q.-F. Yang, Q.-X. Ji, H.~Wang, L.~Wu, B.~Shen, J.~Liu, G.~Huang,
  L.~Chang, W.~Xie, {\em et~al.}, ``Probing material absorption and optical
  nonlinearity of integrated photonic materials,'' {\em arXiv preprint
  arXiv:2111.00105}, 2021.

\bibitem{lu2020toward}
J.~Lu, M.~Li, C.-L. Zou, A.~Al~Sayem, and H.~X. Tang, ``Toward 1\%
  single-photon anharmonicity with periodically poled lithium niobate microring
  resonators,'' {\em Optica}, vol.~7, no.~12, pp.~1654--1659, 2020.

\bibitem{Phillips:11}
C.~R. Phillips, C.~Langrock, J.~S. Pelc, M.~M. Fejer, I.~Hartl, and M.~E.
  Fermann, ``Supercontinuum generation in quasi-phasematched waveguides,'' {\em
  Opt. Express}, vol.~19, pp.~18754--18773, Sep 2011.

\bibitem{Conti:02}
C.~Conti, S.~Trillo, P.~D. Trapani, J.~Kilius, A.~Bramati, S.~Minardi,
  W.~Chinaglia, and G.~Valiulis, ``Effective lensing effects in parametric
  frequency conversion,'' {\em J. Opt. Soc. Am. B}, vol.~19, pp.~852--859, Apr
  2002.

\bibitem{gorodetksy2010determination}
M.~Gorodetksy, A.~Schliesser, G.~Anetsberger, S.~Deleglise, and T.~J.
  Kippenberg, ``Determination of the vacuum optomechanical coupling rate using
  frequency noise calibration,'' {\em Optics express}, vol.~18, no.~22,
  pp.~23236--23246, 2010.

\end{thebibliography}
